\documentclass[%
prl,
rsi,%
amsmath,amssymb,
 reprint,%
]{revtex4-1}

\usepackage{graphicx}
\usepackage{dcolumn}
\usepackage{bm}

\usepackage{blindtext}

\usepackage{epstopdf}

\usepackage{epstopdf}

\usepackage{dcolumn}

\usepackage{bbold}
\usepackage{gensymb}

\usepackage{algorithm}
\usepackage{algpseudocode}

\usepackage{hyperref}

\newcommand{\um}[1]{{\"#1}}

\newcommand{\Rom}[1]{%
  \textup{\uppercase\expandafter{\romannumeral#1}}%
}

\usepackage{color}

\begin{document}
\title{Star-Convex structures as prototypes of Lagrangian coherent structures}
\author{Benedict L\um{u}nsmann}
\affiliation{Max Planck Institute for the Physics of Complex Systems (MPIPKS), 01187 Dresden, Germany}

\author{Holger Kantz}
\affiliation{Max Planck Institute for the Physics of Complex Systems (MPIPKS), 01187 Dresden, Germany}


\begin{abstract}

Oceanic surface flows are dominated by finite-time Lagrangian coherent structures
that separate regions of qualitatively different dynamical behavior.
Among these, eddy boundaries are of particular interest.
Their exact identification is crucial for the study of
oceanic transport processes and the investigation of impacts on
marine life and the climate.
Here, we present a novel method purely based on convexity, a condition that is
intuitive and well-established, yet not fully explored.
We discuss the underlying theory, derive an algorithm that yields comprehensible
results and illustrate the presented method by identifying coherent structures
and filaments in simulations and real oceanic velocity fields.

\end{abstract}

\maketitle

Hydrodynamic mesoscale structures separate the oceans into regions of
qualitatively different dynamical behavior.
These jets, fronts and eddies give rise to short-lived order in the oceans'
upper layer before they inevitably have to pass away in the face of
ever-changing turbulence.
During their lifetime, they have a significant effect on the distribution of
hydrodynamic scalar fields like temperature, oxygen concentration and salinity
\cite{Aristegui1997,Martin2003,Beal2011,Dong2014,Karstensen2015},
as well as nutrient concentration \cite{Martin2003,McGillicuddy2016}
and thereby impact marine life in complex ways
\cite{Martin2003,Karstensen2015,Prants2012,DOvidio2010,DOvidio2013, McGillicuddy2016}.
Moreover, these structures are considered to have a lasting
effect on the climate by providing focused transport of heat and salt over
larger distances \cite{Beal2011}.

Especially eddies, coherently rotating water masses, have gained a
lot of attention in recent years due to their ability to effectively trap water
in their interior.
The trapped water forms a coherent core that does not mix with the ambient water
for a significant amount of time. 
Water in this core may then be coherently transported over larger distances.
In addition, the joint rotation of the enclosed water has the potential to
change the interior nutrient concentration by inducing vertical velocity
fields \cite{Martin2003,McGillicuddy2016}.
This way, eddies can transport warm saline water across the South Atlantic
\cite{Richardson2007,Beron-Vera2013,Froyland2015a}
and have a significant impact
on plankton production
\cite{Bracco2000a,Martin2003,Sandulescu2007,Gaube2014,McGillicuddy2016}.

There exists a variety of different methods that aim to detect eddies and
estimate the boundaries that confine their coherent inner cores.
Generally, such a method is either Eulerian, i.e. it works on velocity field
snapshots, or it is Lagrangian, i.e. it operates on fluid element trajectories.
Eulerian methods include the traditional Okubo-Weiss
criterion \cite{Okubo1971,Weiss1991} but also more recent approaches like
\cite{Isern-Fontanet2003,Chaigneau2008,Itoh2010,Chelton2011,Nencioli2010,Gaube2014}.
The most popular Lagrangian methods can be classified in heuristic
approaches
\cite{Mendoza2010,Mancho2011,Rypina2011,Froyland2015,Hadjighasem2015,Vortmeyer-Kley2016},
probabilistic approaches
\cite{Froyland2013,Ma2013,Froyland2014,Lunsmann2019}
and geometric approaches
\cite{Boffetta2001,Shadden2005,Haller2015,Haller2016}
(for a review of Lagrangian approaches see \cite{Hadjighasem2017}).

Consequently, each method usually comes with its own definition of what it
considers to be an eddy core.
Having many different definitions of coherent structures generally obscures
the interpretation and comparison of results.
However, while employing different concepts and coming to different conclusions,
most approaches agree that eddy cores are mesoscale structures that do not generate
filaments under advection.
Some approaches address this idea by introducing convexity as an indicator
\cite{Hadjighasem2015,Beron-Vera2018}
or as an explicit condition for their eddy core
boundaries \cite{Haller2016,Vortmeyer-Kley2016}.
And it is certainly true that any typical volume that remains convex under
advection does not mix with the surrounding volume.

In this article, we aim to employ convexity as the sole condition for coherent
elliptic structures.
Moreover, we will use this concept to derive an algorithm, the material
star-convex structure search (MSCS-search), that identifies such eddies on the
basis of star-convexity by simply removing non-convex sub-volumes.
\newline

Accepting the idea of convexity as a prototype of coherence for transported
volumes, our following considerations revolve around its formalization and
utilization.

The challenge is to find a way to construct volumes that stay convex under
advection within a predefined time window $[0,\tau]\subset\mathbb{R}$.
Here, advection refers to time evolution under a continuous time-dependent
reversible flow $\Phi: X \times \mathbb{R}\times \mathbb{R}\rightarrow X$ which
maps a volume $S\subseteq X=\mathbb{R}^2$ at time $t$ a time step $\Delta \tau$
into the  future by $\Phi(S,t,\Delta\tau)$.
The size of any volume $S$ is described by its measure
$\mu(S)$ and might change under advection.

We will call any time-dependent volume
$S(t)\subseteq\mathbb{R}^2$ a \emph{structure},
where two classes are of particular relevance for us:
\emph{Material structures} of the form $S(t) := \Phi(S_0, t_0, t-t_0)$ that are
defined by an initial volume $S_0$ transported by the flow $\Phi$ as well as
\emph{convex structures} that are convex for all $t\in[0,\tau]$.

For a structure $S(t)$, we define its maximal material structure
$M_\Phi(S(t))$ as the largest material structure inside $S(t)$.
Likewise, we define the maximal convex structure $C(S(t))$ to be the largest convex
\mbox{structure inside $S(t)$}.

Using this terminology, we can respecify our goal as follows:
Given a flow $\Phi$, a time window $[0,\tau]$, and a structure $S(t)$, we want
to construct the maximal convex material structure $Z(t)\subseteq S(t)$. 
Notably, such a structure has the property $Z(t)=M_\Phi(C(Z(t)))$.
Any structure $S'(t)\subseteq S(t)$ that contains $Z(t)$ is either
non-convex or not a material structure.

Ideally, the concepts of materiality and convexity could be decoupled
such that alternating between a search for maximal material and maximal convex
structures would lead to the correct solution in an iterated fashion.
However, this is not the case here.
Indeed, the search for convex structures might reject regions of the volume
that are necessary to maintain material integrity, not least because the
partitioning of non-convex structures into convex structures is not unique.
This can even lead to maximal convex structures that do not contain any
material structure at all.

For this reason, we propose to relax the problem.
Instead of searching for convex structures, we suggest to search for structures
that remain star-convex with respect to a trajectory $p(t)$.
Using such a trajectory $p(t)$, we are able to define a maximal star-convex
structure $C^\star(S(t),p(t))$ as the largest structure inside $S(t)$ that is
star-convex with respect to $p(t)$.
This structure is unique.

Now, given a time interval $[0,\tau]$, a structure $S(t)$  and a trajectory
$p(t)\subseteq S(t)$ , we search for the maximal star-convex material
structure \mbox{$Z(t)\subseteq S(t)$}.
This is the largest structure in $S(t)$ that fulfills \mbox{$Z(t) =
  M_\Phi(C^\star(Z(t)), p(t))$} and it is an upper bound for the largest convex
material structure in $S(t)$ that contains $p(t)$.
Here, the structures $p(t)$ and $S(t)$ serve as minimal and maximal estimates
of $Z(t)$, i.e. \mbox{$p(t)\subseteq Z(t)\subseteq S(t)$}.

\begin{figure}[h]
  \centering
  \includegraphics[scale=1]{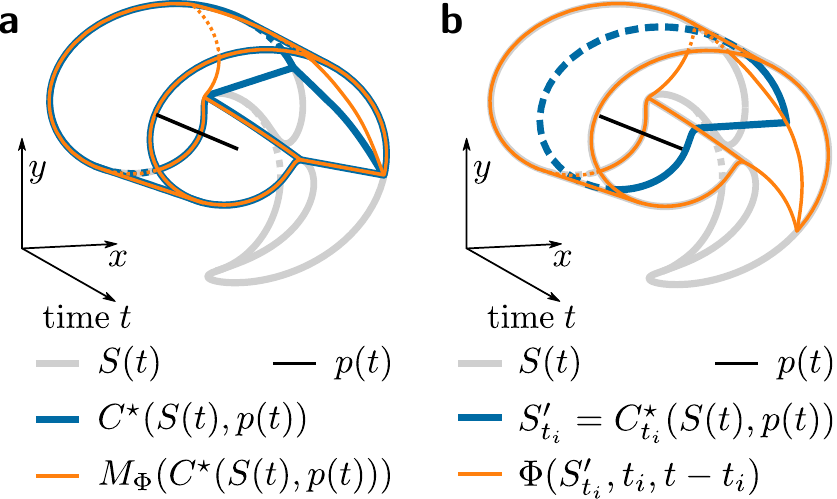}
  \caption{{\bf Two sequences that converge to the maximal star-convex
      material structure $Z(t)$}
    (a) One iteration step and the final result of sequence~(\ref{eq:seq1})
    which converges fast but is computationally costly.
    (b) One iteration step of sequence~(\ref{eq:seq2}) which converges slower
    but is computationally feasible.
   }
  \label{fig:compsketch}
\end{figure}

An iterative solution to this problem is
the following (see Fig.~\ref{fig:compsketch}):
Starting with the structure $S(t)=S^1(t)$, we define the sequence
\begin{align}
  S^{i+1}(t)=M_\Phi(C^\star(S^i(t),p(t))) \label{eq:seq1}\;.
\end{align}
This sequence defines a hierarchy $S^1(t)\supseteq S^2(t)\supseteq \ldots $ with
a trivial lower bound \mbox{$p(t) = M_\Phi(C^\star(p(t),p(t))) \subseteq S^i(t)$}.
Moreover, as neither $C^\star(\cdot, p(t))$ nor $M_\Phi(\cdot)$ remove parts
of the structure $Z(t)$, this sequence converges to
$\lim\limits_{i\rightarrow\infty} S^i(t) = Z(t)$.

However, the maximal material structures $M_\Phi(S'(t))$ inside a non-material
structure $S'(t)$ is quite difficult to find.
In principle, many trajectories that start inside $S'(t)$ have to be
computed to decide which parts leave the structure under advection.
For this reason, evaluation of the sequence~(\ref{eq:seq1}) is computationally
unpractical.
In contrast, maximal star-convex structures can easily be computed by removing
non-star-convex sub-volumes for each time $t\in[0,\tau]$ independently.
This is why we propose to never leave the space of material structures in the
first place by only enforcing star-convexity for single time points (see
Fig.~\ref{fig:compsketch}).

First, we start with an initial material structure $S(t)$ defined by its
volume \mbox{$S_{t_1} = S(t_1)$} at \mbox{time $t=t_1$}.
We compute the maximal star-convex volume
\mbox{$S^1_{t_1} = C^\star_{t_1}(S(t), p(t))$} by removing everything that is not
star-convex at \mbox{time $t=t_1$}.
This volume defines a new structure $S^1(t)$ which is material for all
$t\in[0,\tau]$ and star-convex at $t=t_1$.
Now, we transport this volume using $\Phi$ and see if it remains star-convex.
If it does for all $t\in[0,\tau]$, we found our maximal star-convex material
structure.
If not, we define a new structure $S^2(t)$ by again reducing the volume of
$S^1(t)$ at some time $t=t_2$ to its maximal star-convex volume and try again.
This way, we define the sequence of structures
\begin{align}
  \begin{split}
  S^i(t) &= \Phi(S^i_{t_i}, t_i, t-t_i)\quad\textrm{with}\\
  S^{i+1}_{t_{i+1}} &= C^\star_{t_{i+1}}(S^i(t), p(t))\;.
  \end{split}\label{eq:seq2}
\end{align}
This sequence has the same limit as (\ref{eq:seq1}) but is
computationally less demanding.
\newline

We have seen that the concept of maximal material and maximal star-convex
structures leads to an iterative principle for the construction of maximal
star-convex material structures $Z(t)$.
The only needed parameters are the time interval $[0,\tau]$ and the maximal and
minimal estimates $S(t)$ and $p(t)$ represented by there values at $t=0$.
All that remains is to specify an explicit algorithm for the
sequence~(\ref{eq:seq2}) that generates the limiting structure $Z(t)$.

Computationally, it is useful to describe a volume $S_t$ by its
boundary $\hat{S}_t$, a so-called material line.
These boundaries $\hat{S}_t$ will be represented as
polygons which can be efficiently transported using the flow $\Phi$ although their
number of vertices has to be adjusted regularly to ensure correct approximation
of the enclosed volumes.
In addition, we introduce the number of time steps $N$, the maximal number of
computational cycles $M$, the convergence tolerance $\epsilon$, the maximal
vertex distance $\delta$, and the minimal volume $A$.
However, these parameters only control the numerical stability and the quality
of the results and do not impact the algorithm in any other way.

\begin{figure}[h]
  \centering
  \includegraphics[scale=1]{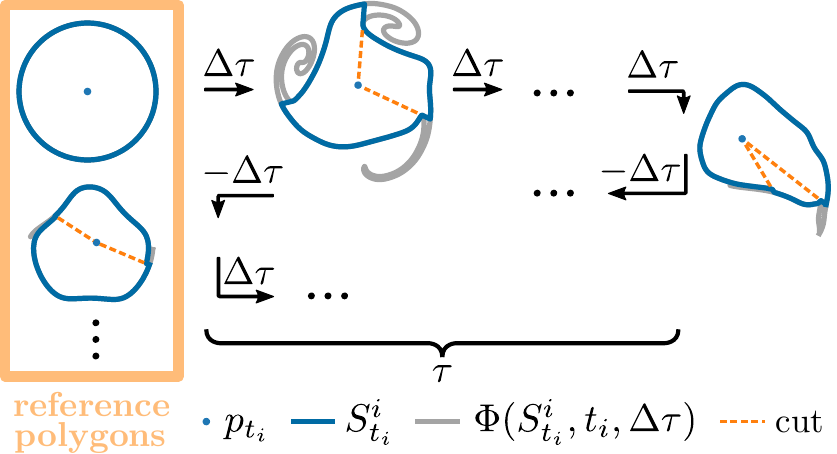}
  \caption{
    { \bf Cutting filaments reveals coherent structures} (color online)
    Starting with an initial polygon $S^1_{t_1}$ as maximal estimate and a point
    $p_{t_1}$ as a minimal estimate of the star-convex material structure (top left), we
    successively integrate the material line by $\Delta \tau$ until the end of
    the desired time-interval $\tau$ is reached (top right). Then we continue by integrating backwards.
    After each integration we enforce star-convexity of the image
    $\Phi(S_{t_i}^i,t_i, t_{i+1}-t_i)$ with respect to $p_{t_{i+1}}$ and
    generate a new material line $S_{t_{i+1}}^{i+1}$.
    If changes between polygons at a reference times point fall below a critical
    values we stop the integration.}
  \label{fig:sketch}
\end{figure}

This finally concludes in the MSCS-search (see
Fig.~\ref{fig:sketch}):

We start with the time interval $[0,\tau]$, an initial material line
$\hat{S}^1_0$ and an initial position $p_0\in\mathbb{R}^2$ as maximal and
minimal estimates for the structure $Z(t)$ at \mbox{time $t=0$}.
We separate the interval in $N$ chunks of length $\Delta \tau$.
Then, we successively transport $\hat{S}_t^i$ and $p_t$ into the future,
reduce $\hat{S}_t^i$ to its maximal star-convex volume and fill the boundaries
with vertices according to the maximal distance between successive vertices
$\delta$.
We do this until we reach the end of the interval \mbox{$t=\tau$} and continue by
integrating backwards in time until we complete a full cycle.

If after several such cycles, the material line at some reference time point
$t_\textrm{ref}$ converges, i.e. if
\mbox{$\frac{\mu(S_{t_\textrm{ref}}^\textrm{ref})-\mu(S_{t_\textrm{ref}}^i)}{\mu(S_{t_\textrm{ref}}^\textrm{ref})}<\epsilon$},
we found our solution $Z(t)$.
Otherwise, if the area becomes too small \mbox{$\mu(S_{t_\textrm{ref}}^i)<A$} only the trivial solution
$Z(t)=p(t)$ remains.
If more than $M$ cycles are needed, the structure converges too slowly.
 
The initial material line $\hat{S}_0^1$ should be chosen generously and as
smooth as possible in order to avoid numerical complications.
The initial position $p_0$ can be well estimated using simple proxies (see
Results).
\newline

In order to display the potential of our approach, we use the MSCS-search to
identify coherent structures in artificial and empirical velocity fields.
Moreover, we demonstrate that by changing the observation horizon $\tau$
it is possible to investigate when which parts of the structure are detached or
entrained as filaments.

\begin{figure}[h]
  \centering
  \includegraphics[scale=1]{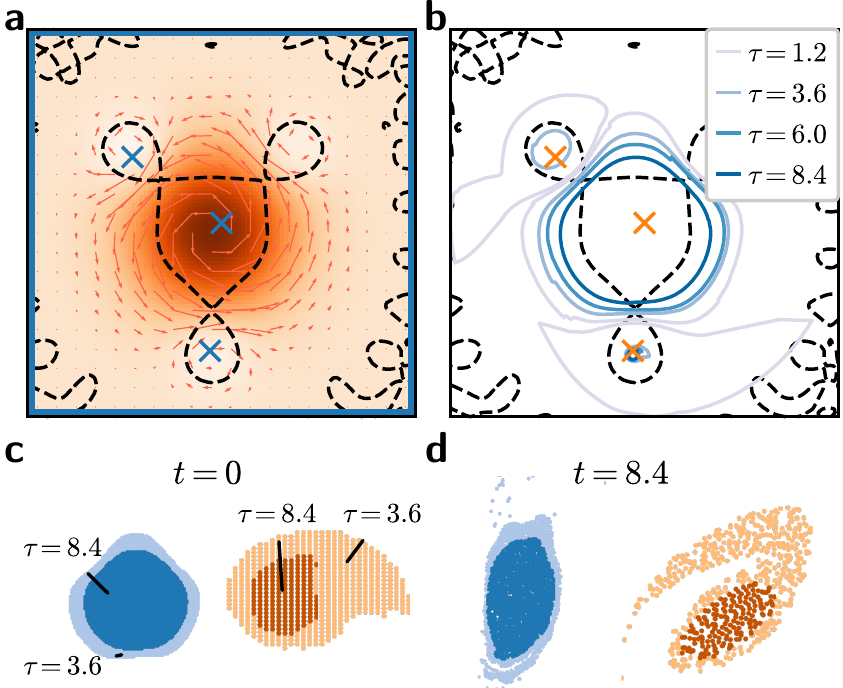}
  \caption{
    {\bf Convex structures for different integration times $\tau$} (color
    online)
    (a) Model initialized with large central vortex and three surrounding
    vortices of different strength. Stream function in the background, dashed
    lines indicate zero level of Okubo-Weiss criterion. We analyzed three
    vortices marked by blue crosses which denote the minimal estimates.
    The blue rectangle denotes the maximal estimate for each structure.
    (b) Results of analysis for different \mbox{integration times $\tau$}. Medium and
    larger integration times converge to similar structures of reasonable
    proportions.
    Results of the upper left vortex indicate disintegration of any convex
    structure for larger integration times.
    (c) Initial positions ($t=0$, left) and final positions ($t=8.4$, right) of
    test tracers started within the boundaries of the central (blue) and
    lower (orange) structure for different integration times (light:
    $\tau=3.6$, dark: $\tau=8.4$).
    While the boundaries for medium integration times generate filaments,
    larger integration times result in robust structures.
  }
  \label{fig:piceuler}
\end{figure}

First, we apply our method to flow fields generated by the two-dimensional Euler
equation on a square domain with periodic boundary conditions.
The two dimensional inviscid flow is initialized with a central large and
strong vortex that is surrounded by three smaller and weaker eddies of opposite
sense of rotation (see Fig.~\ref{fig:piceuler}a).

We want to determine the coherent cores of three eddies: the central
strong eddy, the smaller eddy in the upper-left corner and the smaller eddy at the
bottom.
For this reason, we compute the \mbox{Okubo-Weiss criterion $Q$}
\cite{Okubo1971,Weiss1991} for the initial velocity field and choose positions
$p_{\textrm{ctr},1}$, $p_{\textrm{ul}, 1}$, $p_{\textrm{bot}, 1}$ in the
vicinity of the corresponding minima.
The Okubo-Weiss criterion compares stretching and shear flow with rotation and
is an established but Eulerian proxy for vortex positions.
The positions
$p_{\textrm{ctr},1}$, $p_{\textrm{ul}, 1}$, $p_{\textrm{bot}, 1}$
serve as individual minimal estimates for each eddy.
For all three eddies, we choose the complete domain as the maximal estimate
$\hat{S}_0^1$ (see Fig.~\ref{fig:piceuler}a).

We use the MSCS-search for different observation horizons $\tau$ to compute
estimates for the largest star-convex material structure $Z(t)$ in each eddy (see
Fig.~\ref{fig:piceuler}b).

The results show that the material lines for small integration times $\tau$ are
quite unlikely to correspond to the boundaries of coherent structures.
This was to be expected, since the results returned by the MSCS-search
are inevitably connected to the predefined time window.
In the extreme case of $\tau\rightarrow 0$ the algorithm would simply produce
the star-convex initial boundary $\hat{S}_0^1$.

For larger integration times however, the structures become smaller and
smaller and generate a hierarchy of nested sets.
Again, this is an expected phenomenon since larger time windows will only
generate smaller structures.

The difference between material lines for different integration times
corresponds to filaments that are shed from the eddy core in the time between the
ends of time windows.
Thus, we are able to study the shrinkage of the coherently
transported volume.
Of course, inverting the time direction would enable the investigation of
filament entrainment.

Starting from the eddy in the upper left corner the area enclosed by the
star-convex material becomes too small already after intermediate integration
times $\tau$.
Hence, we conclude that no coherent structure exists for longer time intervals
that include the tracer $p_{\textrm{UL},1}$ released at $t=0$.

The inferred material lines of the large central eddy almost appear to converge
for longer integration times.
This indicates a persistent coherent structure.

In order to test the coherence of the inferred volumes, we release test tracers
within the boundaries of the central and the lower vortex at $t=0$ and compute
their trajectories until the end of the largest time window $t=8.4$ (see
Fig.~\ref{fig:piceuler}c).
We notice that tracer clouds within the boundaries of the star-convex material
line inferred for $\tau=3.6$ generate filaments at $t=8.4$ while tracers within
the boundaries of the material for $\tau=8.4$ do not, just like our approach
predicts.
However, it is intriguing that the generated filaments are rather small and
remain in the vicinity of the more coherent structure.
Thus, even if a volume generates filaments under advection and is thus rejected
by our approach, it might still appear to be coherent if these filaments are not
resolved.

\begin{figure}[h]
  \centering
  \includegraphics[scale=1]{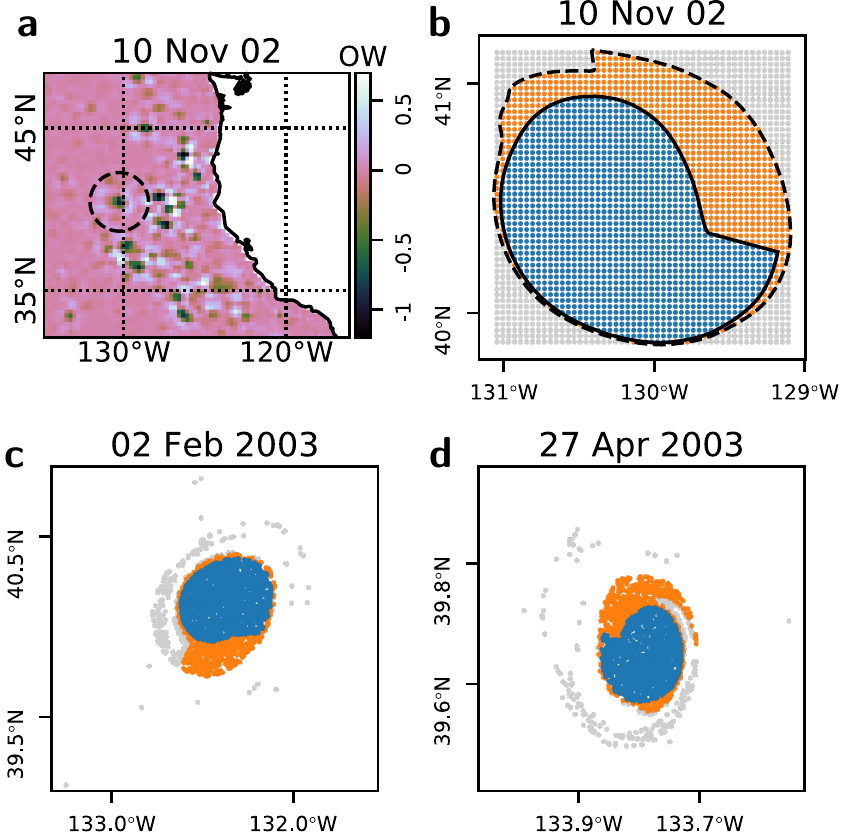}
  \caption{{\bf Star convex material structure in the North Pacific} (color online)
  (a) Okubo-Weiss criterion, time and position of initial structures. 
  (b) Results of MSCS-search for $\tau=12$~weeks (orange) and
  $\tau=24$~weeks (blue) at $t=0$ (c) at $t=12$~weeks and (d) at $t=24$~weeks.
  Gray tracers indicate mixing with ambient water, most of which leave the
  displayed domain.
  }
  \label{fig:scb}
\end{figure}

In addition, we apply the MSCS-search to an open dataset of surface velocities
in the North Pacific with spatial and temporal resolution of $0.25\degree$ and
$1$~day respectively \cite{Risien15} (see Fig.~\ref{fig:scb}).
The initial maximal and minimal structures are again chosen on the basis of the
Okubo-Weiss criterion (see Fig.~\ref{fig:scb}a) and yield reasonable results for
$\tau=12$ and $24$~weeks (see Fig.~\ref{fig:scb}b).
The integration of particles reveals that both structures do not generate
filaments for $t=12$~weeks and do not even disperse significantly for
$t=24$~weeks while the surrounding material generates filaments and mixes with the
ambient water (see Fig.~\ref{fig:scb}c/d).
During transport, the water mass undergoes considerable contraction.

\emph{Numerics.} 
The Euler equation is solved using a spectral ansatz with $11$ Fourier components.
We realize its integration and the integration of particles using a standard
adaptive RK45 scheme.
We choose the time step of
$\Delta t=0.1$, the maximal vertex distance of $\delta=0.01$ and the minimal
polygon area of $A=0.2\cdot 10^{-3}$.
To facilitate the search for larger integration times, we
initialize their initial maximal estimates $\hat{S}_0^1$ as the smallest circular
polygon with the minimal estimate $p_0$ as its center that still contains the results
of smaller integration times.

For the oceanic velocity field, the particle integration was realized using linear
interpolation in time and space, a standard RK45 approach and a maximal time
step of $0.25$~days.
\newline

We argued that convexity is an intuitive condition for coherent structures that
many established methods could agree on.
On this basis, we have used the concept of star-convexity to derive an iterative
principle for the estimation of specific volumes that remain star-convex under
advection with a flow $\Phi$.

Our MSCS-search algorithm, exploiting this principle, requires little prior
knowledge or parameter tuning. It yields convincing results as we demonstrated
for both artificial and empirical time dependent velocity fields.

The fixed time window $[0,\tau]$ required by our approach enables us to detect
finite-time coherent structures and to study eddy decay and filament entrainment
in detail.
Additionally, the approach is objective in the sense of \cite{Beron-Vera2013} and able to
treat divergent velocity fields as they are typical for surface velocities.
Moreover, it does not depend on auxiliary parameters that must be tweaked or
tuned to generate desired results.
Instead, all additional parameters only control the algorithm's numerical stability.
Their ideal values are known and their impact is self-explanatory.
In conclusion, this approach seems well applicable to real oceanic velocity
fields.

Future extensions could for instance include avoidance of unnecessary repeated
particle integrations and the enabling of parallel computation.

%

\end{document}